\documentclass[12pt]{article}
\usepackage{times}
\usepackage{geometry,wrapfig,setspace}
\usepackage[utf8]{inputenc}
\usepackage{enumitem,amssymb,graphicx}
\usepackage{ragged2e}
\newlist{thematic}{itemize}{8}
\setlist[thematic]{label=$\square$}
\usepackage{pifont}
\newcommand{\um}{$\mu$m}
\newcommand{\etal}{et.~al.}
\newcommand{\sfr}{M$_{\odot}$\,yr$^{-1}$}

\def\ltsima{$\; \buildrel < \over \sim \;$}
\def\simlt{\lower.5ex\hbox{\ltsima}}
\def\gtsima{$\; \buildrel > \over \sim \;$}
\def\simgt{\lower.5ex\hbox{\gtsima}}

\begin{document}
\raggedright
\huge
Astro2020 Science White Paper \linebreak

Taking Census of Massive, Star-Forming Galaxies formed $<$1\,Gyr After the Big Bang\linebreak
\normalsize

\noindent \textbf{Thematic Areas:} \hspace*{60pt} 

\makebox[0pt][l]{$\square$}\raisebox{.15ex}{\hspace{0.1em}$\checkmark$}    Galaxy Evolution   
\linebreak  

\vspace{-3mm}
\textbf{Principal Author:}

Name:	Caitlin M. Casey
 \linebreak						
Institution:  The University of Texas at Austin
 \linebreak
Email: cmcasey@utexas.edu
 \linebreak
Phone:  +1(512) 471-3405
 \linebreak
 
\vspace{-3mm}
\begin{spacing}{0.8}
  {\small
    \textbf{Co-authors:} 
    Peter Capak (IPAC/Caltech),
    Johannes Staguhn (NASA Goddard/Johns Hopkins University),
    Lee Armus (IPAC/Caltech),
    Andrew Blain (University of Leicester),
    Matthieu Bethermin (LAM),
    Jaclyn Champagne (University of Texas at Austin),
    Asantha Cooray (University of California Irvine),
    Kristen Coppin (University of Hertfordshire),
    Patrick Drew (University of Texas at Austin),
    Eli Dwek (NASA/Goddard),
    Steven Finkelstein (University of Texas at Austin),
    Maximilien Franco (University of Hertfordshire),
    James Geach (University of Hertfordshire),
    Jacqueline Hodge (Leiden Observatory),
    Jeyhan Kartaltepe (Rochester Institute of Technology),
    Maciej Koprowski (University of Hertfordshire),
    Claudia Lagos (International Centre for Radio Astronomy Research, University of Western Australia),
    Desika Narayanan (University of Florida),
    Alexandra Pope (University of Massachusetts Amherst),
    David Sanders (University of Hawai'i),
    Irene Shivaei (University of Arizona),
    Sune Toft (DAWN/University of Copenhagen),
    Joaquin Vieira (University of Illinois),
    Fabian Walter (Max Planck Institute for Astronomy),
    Kate Whitaker (University of Connecticut),
    Min Yun (University of Massachusetts Amherst),
    Jorge Zavala (University of Texas at Austin).
  }
\end{spacing}

\justify \textbf{Abstract:} Two decades of effort have been
poured into both single-dish and interferometric millimeter-wave
surveys of the sky to infer the volume density of dusty star-forming
galaxies (DSFGs, with SFR$\simgt$100\sfr) over cosmic time.  Though
obscured galaxies dominate cosmic star-formation near its peak at
$z\sim2$, the contribution of such heavily obscured galaxies to cosmic
star-formation is unknown beyond $z\sim2.5$ in contrast to the
well-studied population of Lyman-break galaxies (LBGs) studied through
deep, space- and ground-based pencil beam surveys in the
near-infrared.
Unlocking the volume density of DSFGs beyond $z>3$, particularly
within the first 1\,Gyr after the Big Bang is critical to resolving
key open questions about early Universe galaxy formation:
{\bf (1)} What is the integrated star-formation rate density of the Universe
in the first few Gyr and how is it distributed among low-mass
galaxies (e.g. Lyman-break galaxies) and high-mass galaxies
(e.g. DSFGs and quasar host galaxies)?
{\bf (2)} How and where do the first massive galaxies assemble?
{\bf (3)} What can the most extreme DSFGs teach us about the mechanisms of
dust production (e.g. supernovae, AGB stars, grain growth in the ISM)
$<$1\,Gyr after the Big Bang?
We summarize the types of observations needed in the next decade to
address these questions.

\pagebreak

\noindent {\bf Background:} Among all star-forming galaxies, DSFGs are
the most massive and extreme: they are characterized by high star
formation rates, typically above 100\,M$_\odot$\,yr$^{-1}$, and
typical stellar masses above 10$^{10}$\,$M_\odot$. As a byproduct of
their high star-formation rates they are very dusty galaxies, whereby
$>95$\%\ of the emission from hot, young stars is obscured by dust and
reprocessed from rest-frame UV to rest-frame far-infrared emission
(see reviews of Sanders \&\ Mirabel 1996, Blain \etal\ 2002, and Casey, Narayanan \&\ Cooray 2014).
Today we know that, though rare in the nearby Universe, these DSFGs
are 1000$\times$ more common at $z\sim2$, such that they dominate all
of cosmic star formation (e.g. Gruppioni \etal\ 2013).  However, their
volume density at $z\simgt2.5$ is still unconstrained (see Figure~1).
Completing an accurate census predicates any representative studies of their
physical evolution and fundamental role in driving galaxy evolution.

Unfortunately the strategy which has been so successful for studying
the rest-frame optical and near-infrared emission from `normal'
high-redshift galaxies (e.g. Lyman-break galaxies, LBGs)---deep pencil
beam surveys in extragalactic legacy fields---cannot (and has not)
enabled a clear census of dust-obscured galaxies at $z>3$ (e.g. Dunlop
\etal\ 2016, Franco \etal\ 2018).  This is because DSFGs are
relatively rare compared to LBGs (at $z\sim2$ there are roughly
100\,LBGs for every DSFG, despite contributing equal amounts to cosmic
star-formation); this is a direct result of DSFGs being at the tip of
the galaxy mass function.  Whitaker \etal\ (2017) demonstrate directly
that, indeed, it is galaxies with the highest stellar masses
($>$10$^{10}$\,M$_\odot$) that are most obscured by dust ($>$90\%),
while fainter, low-mass galaxies have lower total obscuration
fractions ($\sim$20-80\%).  This trend with mass results in very
different forms of the galaxy luminosity function (LF) in the
rest-frame UV vs. the rest-frame FIR: the UVLF is characterized by a
steep faint-end slope and an exponential fall-off at the bright end
(e.g. Finkelstein 2016), while the IRLF is characterized by a shallow
faint-end slope and broken powerlaw with increasing luminosity.

Thus, the vast majority of DSFGs identified to date have been found in
wide-area ($>$0.1--100\,deg$^2$) surveys from single-dish FIR and
millimeter telescopes like JCMT/SCUBA and SCUBA-2 (starting with Smail
\etal\ 1997), {\it Herschel Space Observatory} (e.g. Elbaz
\etal\ 2011, Casey \etal\ 2012b,c), and the AzTEC instrument on JCMT,
ASTE and the LMT (e.g. Scott \etal\ 2008, Aretxaga \etal\ 2011).
Meticulous multi-wavelength follow-up of $\sim$100s of these sources
over the past decades have revealed that the vast majority of them sit
at redshifts $1\simlt z\simlt 3$ (e.g. Chapman \etal\ 2005, Cowie
\etal\ 2018) and are either fueled by major galaxy mergers or
represent a class of particularly gas-rich massive disk galaxies
undergoing intense star formation (e.g. Ivison \etal\ 2012, Hodge
\etal\ 2013, 2016).

The primary hindrance to the identification and characterization of
DSFGs at {\it earlier} epochs comes from the difficulty in accurately
identifying the handful that might sit at higher redshifts.  This
difficulty is, in part, due to the very negative K-correction at
(sub)millimeter wavelengths: though it is a `blessing' that DSFGs are
just as bright at $z\sim10$ as at $z\sim1$, it substantially
obfuscates our ability to robustly identify their redshifts with
long-wavelength emission alone.  (Sub)millimeter colors can trace
either a DSFGs' dust temperature or (rough) redshift, a degeneracy
that hinders the quick characterization of DSFGs.  The large beamsizes
of typical DSFG selection surveys further obfuscates their accurate
identification.

As a thought experiment (one presented in Casey \etal\ 2018a and in
Figure~1), if we are to distinguish between two extremely
different star-formation rate densities (SFRDs) to describe the
possible high-$z$ volume density of DSFGs, we would need to accurately
identify the redshifts of {\it all} ($>$98\%) DSFGs selected at, e.g.,
850\um.  This conclusion is reached by a forward evolution model,
asserting an adopted IRLF and analyzing how submm number counts and
sample redshift distributions would present.
Out of a sample of 100 DSFGs, the extreme `dust rich' model in Fig.~1
predicts 7 DSFGs to sit above $z>4$ while the 
`dust poor' model predicts only 3 DSFGs to sit above $z>4$ (also see
Casey \etal\ 2018b for implications with ALMA deep fields).
Unfortunately, the most spectroscopically-complete samples of DSFGs
(like the ALESS and SPT samples; Vieira \etal\ 2013, Danielson
\etal\ 2017) are, at most, $\sim$50-90\%\ complete, thus lacking the
precision necessary to draw accurate conclusions as to the number
density of early-Universe DSFGs.

This concept paper focuses on strategies for finding and
characterizing such early Universe ($z>3$) DSFGs, with particular
focus on pushing toward higher redshifts when DSFG formation scenarios
place more stringent constraints on the formation of the earliest
massive galaxies, $<$1\,Gyr after the Big Bang.

\begin{figure}
\centering
\vspace{-5mm}
\includegraphics[width=0.85\textwidth]{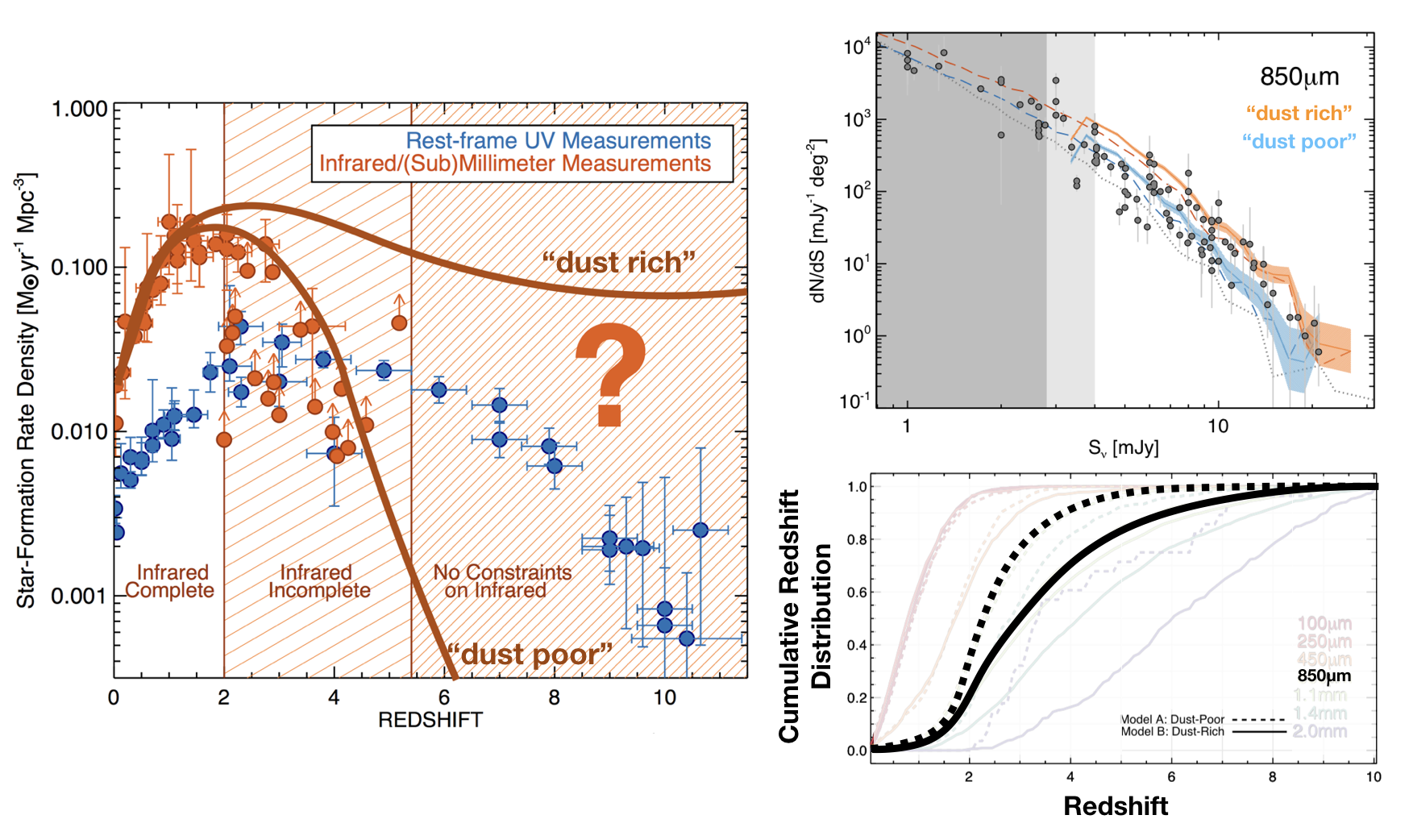}
\vspace{-5mm}
\begin{spacing}{0.8}
\caption{\small {\it (Figures modified from Casey \etal\ 2018a.)}
  {\bf Left:} The cosmic star-formation history of the Universe as
  measured at rest-frame UV wavelengths (blue points) and infrared
  through millimeter measurements (orange points). No dust
  correction has been applied to the UV.
  While rest-frame UV measurements now reach out beyond $z>10$ thanks
  to deep {\it HST} near-infrared imaging campaigns, in contrast
  surveys of obscured emission in galaxies is incomplete at $z>3$ and
  completely unconstrained at $z>5$.  Casey \etal\ (2018a) tests two
  dramatically different possibilities for the obscured fraction of
  the SFRD, shown by the dark orange lines. They show (shockingly)
  that we do not yet have data good enough to distinguish between
  these two models. {\bf Upper right:} The difference between these
  two extreme `dust poor' and `dust rich' models in 850\um\ number
  counts against data (gray points). {\bf Lower Right:} Difference in
  the expected redshift distribution for 850\um-selected DSFGs for the
  two models: no dataset has the precision needed to distinguish even
  between these two models.}
\end{spacing}
\end{figure}

\vspace{2mm}
\noindent {\bf Broader Importance to Galaxy Evolution:}
Knowing the prevalence of dust-obscured star formation is particularly
important at $z\simgt 5$, when cosmic time becomes a constraint on the
physical processes involved in producing dust, metals and stars seen
in galaxies.  For example, if DSFGs contribute significantly to cosmic
star-formation at this epoch (as suggested by new observational
results from Zavala \etal\ 2018), then it would imply that dust
production mechanisms must be particularly efficient and likely happen
via supernovae (Matsuura \etal\ 2011, Dwek \etal\ 2014), combined with
low destruction rates, rather than from coagulation in the upper
atmospheric winds of AGB stars, or coagulation or accretion of dust in
the ISM (Matsuura \etal\ 2006, 2009, Jones \etal\ 2013).  It might
also have a fundamentally different composition (e.g. De Rossi
\etal\ 2018).

Furthermore, not only do early DSFGs teach us about dust formation,
but they also shed light on the formation of the first massive halos.
Their halos are just as massive, if not more massive than quasar host
galaxies at similar epochs (e.g. Wang \etal\ 2011, Bethermin
\etal\ 2013, Maniyar \etal\ 2018), and yet the number density of DSFGs
may be as much as 10-100$\times$ higher than rare quasars given the
prolonged duration of a starburst episode with respect to the
short-lived quasar phase.
The detection of DSFGs like SPT0311 at $z=6.9$ (Strandet \etal\ 2017,
Marrone \etal\ 2018) with a measured gas mass $>$10$^{11}$\,M$_\odot$
push the limits: at this epoch, the most massive halo detectable in
the entire observable universe would only be $\sim$3-5$\times$ more
massive than what is measured, and the implied baryon collapse
efficiency near 100\%.
Are there more similarly-extreme DSFGs at these epochs, or at higher
redshifts, to be found?  Without a measurement of the underlying
number density of DSFGs around this epoch, we lack a good handle on
the relative rarity of massive halos like SPT0311, thus few {\it
  observational} constraints on hierarchical formation itself during
the first Gyr.

\vspace{2mm}
\noindent {\bf Current Barriers to Progress:}
There are several reasons the aggregate $z>3$ DSFG population has
proved elusive, most of which are directly attributable to a dearth of
data or necessary {\it complete} follow-up on existing targets.
Overcoming these barriers to progress requires a careful analysis of
their causes (many are discussed at further length in Casey
\etal\ 2018a,b, and Zavala \etal\ 2018).  Here we summarize what we
see as the primary limitations to analysis in the current era:

{\bf 1. DSFGs are rare and the faint-end slope of the IRLF is shallow.}
Deep, pencil beam millimeter-wave surveys on the order of a few tens
of arcmin$^2$ only result in detection of tens of sources, the vast
majority of which sit at $z<3$.  This is what has been found by
pioneering ALMA deep fields conducted at 1\,mm (Dunlop \etal\ 2016,
Aravena \etal\ 2016, Franco \etal\ 2018).  The core reason that DSFGs
are rare is because dust-obscured star formation preferentially lives
in massive galaxies (as found by Whitaker \etal\ 2017).  Because the
sky density of DSFGs is relatively low, large areas of sky need to be
mapped (on the order of $\simgt$1-10\,deg$^2$) to find a sufficient
number of sources ($\simgt$100), of which only a handful (at most)
will sit at $z>4$ (see Casey \etal\ 2018a,b).  Due to the strong
negative K-correction at millimeter wavelengths, pushing deep pencil
beam surveys deeper will, perhaps counter-intuitively, result in
detections of lower redshift ($z\sim1$), less extreme systems (like
LIRGs, with 10$\simlt$SFR$\simlt$100\,M$_\odot$) rather than an
increase in the higher-redshift DSFG population (e.g. Bethermin
\etal\ 2015).  These less extreme systems are likely to be well
characterized by optical/near-infrared surveys.

{\bf 2. DSFGs at $z\simgt 4$ are a needle in a haystack relative to
  those at $1<z<3$, thus easy to mis-identify.}
The expected number counts (density of sources of certain flux
densities) at 850\um--1\,mm {\it and} the anticipated shape of the
redshift distribution is largely \underline{insensitive} (within
uncertainties) to the two dramatically different hypothetical
universes presented in Figure~1.  This is because the vast majority of
all DSFGs sit at $1<z<3$ (e.g. Danielson \etal\ 2017, Cowie
\etal\ 2018), where they are known to dominate all of cosmic star
formation.  Even in the extreme `dust rich' case, whereby DSFGs are
proposed to dominate cosmic star formation at $z>4$, their number
density on the sky would be rather low. This phenomenon is also
present in LBG samples, although redshifts are far more
straightforward to infer for LBGs, and so confusing a $z\sim2$ LBG
with a $z\sim6$ LBG is unlikely, whereas it is very possible to
confuse DSFGs at these two epochs.  This results in a `needle in the
haystack' problem for $z\simgt 4$ DSFGs: they are incredibly rare and
not at all easy to pick out from the average millimeter flux density
selected sample.\\
Further aggravating the `needle in the haystack' problem are DSFG
mis-identifications.  In other words, it is rather easy to
mis-identify a DSFG's redshift or optical/near-IR counterpart,
particularly when DSFGs are originally selected in
poor-spatial-resolution single-dish datasets, with beamsizes
15--30$''$ across, and only some have interferometric follow-up from
ALMA that spatially locates the position of millimeter wave emission.
Even in cases of clear positional localization via interferometry,
DSFGs can be confused with foreground optically-luminous galaxies
within 1--2$''$, and only identified as background sources through
serendipitous means: for example, through detection of millimeter line
emission with ALMA or PdBI/NOEMA (GN20, HDF850.1, and AzTEC-2; Daddi
\etal\ 2009, Walter \etal\ 2012, Jiminez-Andrade \etal, in prep).
These mis-identifications can easily wash out the census of DSFGs at
$z>4$, where precision and sample completeness is needed at the level
of 98--99\%.

{\bf 3. Selection of DSFGs has been carried out primarily at $\lambda\le 1$\,mm.} 
(Sub)millimeter extragalactic surveys have largely been carried out at
850\um\ (SCUBA \&\ SCUBA-2), 250--500\um\ ({\it Herschel}/SPIRE) or
1\,mm (AzTEC/ALMA), with the largest area ($\gg$1\,deg$^2$) covered
only by {\it Herschel}.  As shown in Figure 2, the shorter wavelengths
are less sensitive surveys are to DSFGs at $z>3$.  1\,mm surveys are
sensitive to high-$z$ DSFGs, though samples are dominated ($\sim$95\%)
by the more numerous $z<3$ DSFGs. Selection at longer wavelengths, in
particular 2\,mm, would effectively filter-out the low-$z$ DSFGs while
pushing the sensitivity of surveys deeper at higher-$z$.
Unfortunately, no significant surveys have yet been carried out at
2mm, save small GISMO maps in HDF and COSMOS (Staguhn \etal\ 2014,
Magnelli \etal\ 2019, submitted).

\hspace{-0.4in}
\begin{minipage}{3.4in}
\includegraphics[width=3.4in]{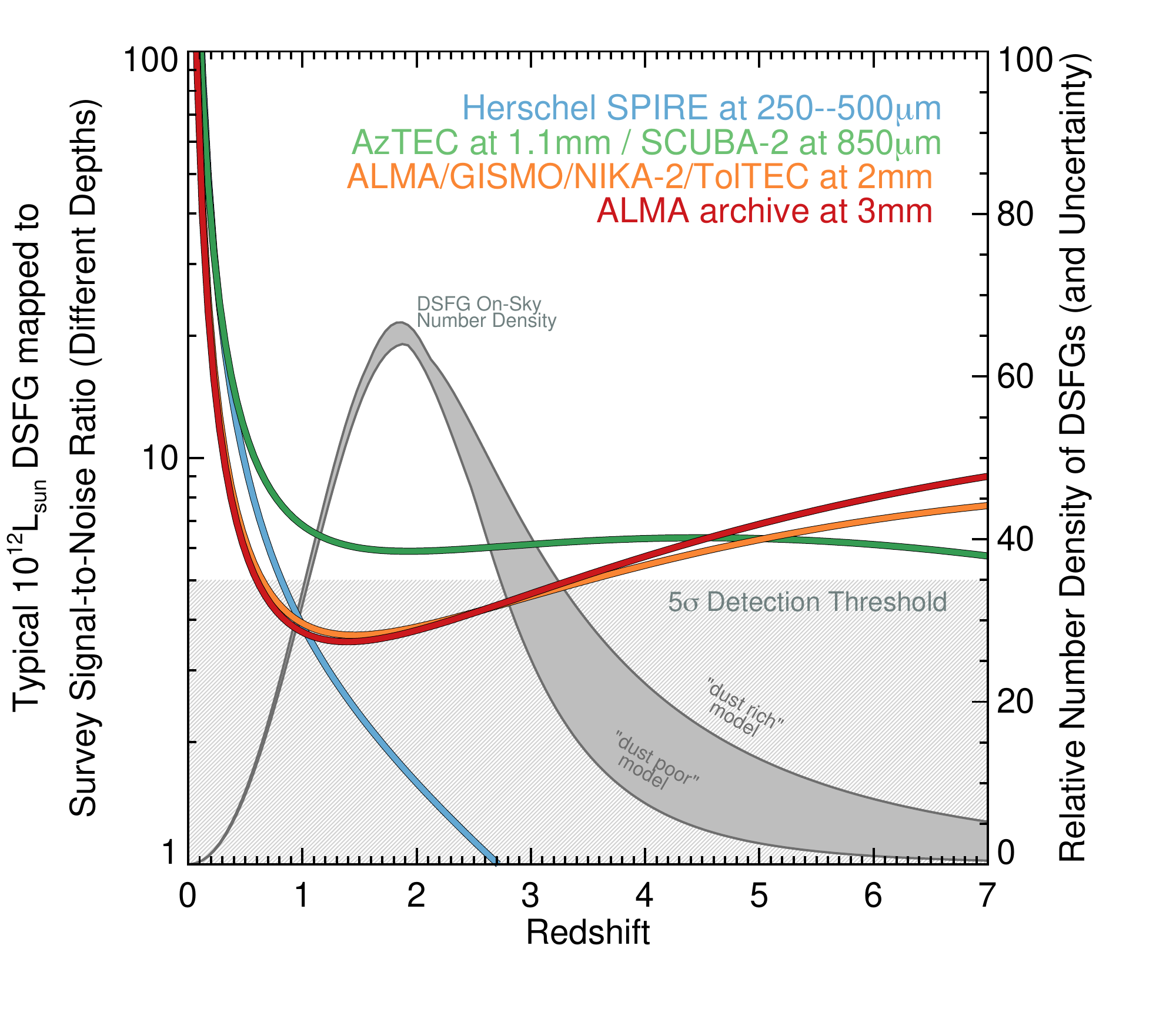}
\end{minipage}
\begin{minipage}{3.2in}
\begin{spacing}{0.8}
{\small Figure 2: The detection signal-to-noise of DSFGs in (sub)mm
  datasets as a function of redshift, scaled by typical
  survey sensitivity for a 10$^{12}$\,L$_\odot$ system.
  Wavelengths shown are 250--500\um\ ({\it Herschel}/SPIRE; blue),
  1.1\,mm (AzTEC; green, also representative of SCUBA-2 at 850\um),
  2\,mm (ALMA/GISMO/NIKA-2/TolTEC, orange), and 3\,mm (ALMA, red, see
  Zavala \etal\ 2018c).  Below a signal-to-noise of five, sources are
  not detectable (light gray region).  On the right axis (and gray
  band) is the relative number density of DSFGs, showing uncertainty
  projected from the dust rich and dust poor models from Figure~1
  (dark gray region).  {\it This figure shows that the vast majority
    of DSFGs sit at $1<z<3$, and the most effective filter for the
    highest-$z$ DSFGs is selection at 2--3\,mm, which greatly
    mitigates the `foreground.'} These tracks do account for CMB
  heating (da Cunha \etal\ 2013) and an average $\lambda_{\rm
    peak}=100$\,\um\ (rest-frame).}
\end{spacing}
\end{minipage}

\vspace{-2mm}
\noindent \underline{\bf Strategy to identify the highest-redshift Obscured Galaxies in the 2020s:}
As we look toward the next generation of instruments and facilities
that could be used to detect highly obscured, massive galaxies like
DSFGs in the early Universe, we recognize that the strategy our
community must adopt will be quite different than the strategy that
was so successful for pioneering {\it Hubble} deep fields.  Drilling
deeper in the millimeter is likely to result in a modest increase in
source density with a lower average redshift for fainter sources.
This is because the obscured galaxy luminosity function is much
shallower at the faint end than the UV luminosity function and due to
the very strong negative K-correction in the millimeter.  Thus, {\bf a
  census of obscured galaxies at $z>3$ requires ambitious large
  surveys ($\simgt$1-10\,deg$^{2}$) carried out at 1.4\,mm--2\,mm with
  large single-dish millimeter observatories with large fields of
  view.  To-date, no such surveys have been carried out that fulfill
  this criterion}; this, despite the fact it was a top priority from
the 2010 decadal review at the mid-scale (the CCAT project).

The US is one of the few regions that does not have open access to the
single-dish (sub)millimeter facilities needed to carry out this work,
despite our tremendous investment in millimeter wave astrophysics
through ALMA\footnote{There is no US access to the IRAM 30\,m, and
  there is institutionally-restricted access in the US to the JCMT
  (the University of Hawai'i), the LMT (University of Massachusetts
  Amherst), and the future, descoped CCAT prime project (Cornell).}.
In order to take an accurate census of the highest-redshift massive
galaxies and successfully pursue a number of other science objectives
that are unique to such facilities, the US community needs to
prioritize either the funding of existing facilities (like JCMT or the
LMT) or build a 30--50\,m facility (like the originally planned CCAT,
or join the new European AtLAST project).
In the near-term, the instrumentation that will contribute to this census are:

\noindent {\bf GISMO:} The 2\,mm GISMO instrument conducted two blank
field mapping projects at the IRAM 30\,m (Staguhn \etal\ 2014,
Magnelli \etal\ 2019) and has the potential to continue cutting-edge
2\,mm mapping in the very near future at a facility like JCMT or LMT.
Timing is crucial for {\it JWST} follow-up on the galaxies' stellar
emission. Selection at 2\,mm is an efficient method of isolating the
highest-redshift DSFGs by filtering out the majority of lower redshift
DSFGs, and GISMO has provided the first efficient way to reach
$<$1\,mJy depths needed to detect a sufficient number of sources per
area on the sky (this is needed because the 2\,mm number counts,
despite being a useful way to isolate high-$z$ sources, are very steep
and shallower maps will detect very few sources).

\noindent {\bf NIKA-2 at the IRAM 30\,m:} NIKA-2 is going to push both
  1\,mm and 2\,mm surveys at the IRAM 30\,m over the next decade.  The
  two bands could work well to isolate the highest redshift DSFGs of
  interest in blank surveys.

\noindent {\bf TolTEC at the LMT:} The TolTEC instrument will be installed
  on the LMT in the early 2020s and carry out ground-breaking
  continuum surveys at 1\,mm, 1.4\,mm, and 2\,mm with unparalleled
  mapping speed using MKID technology.  Similar to NIKA-2, the three
  bands could work to simultaneously and effectively filter DSFGs into
  intermediate and high-redshift subsamples.

\noindent {\bf SPT\,3G:} Though a CMB experiment, the
SPT\,3G survey will be substantially deeper than the original SPT
surveys, forging into parameter space where unlensed DSFGs would be
detectable.  While the telescope is much smaller than other
facilities (10\,m), which results in a larger beamsize, confusion
noise will not dominate as the source density at 2--3\,mm is low. 
 SPT\,3G will cover more sky than TolTEC and more
efficiently detect rarer, brighter DSFGs.

\noindent {\bf ALMA:} ALMA is extremely sensitive and versatile,
despite being fundamentally limited to very narrow regions of sky, for
lack of its wide-area mapping capability.  Nevertheless, it can play a
crucial role in the census of DSFGs at high-$z$ through programs
pushing in new directions.  For example, the first 2\,mm map (deeper
than existing GISMO maps) is in the Cycle 6 ALMA queue from the COSMOS
team, requiring 41 hours in band 4 to cover 230\,arcmin$^2$ to
0.9\,mJy RMS, though it is not yet observed.  The depth and angular
resolution of such deep mosaics can discern between drastically
different models of DSFG number density at $z>4$.  Similarly, existing
3\,mm (band 3) programs focused on alternate science goals
(e.g. detection of CO at $z\sim1-2$) can be sufficiently deep to
reveal serendipitous continuum sources (as in Zavala \etal\ 2018c)
over hundreds of arcmin$^2$.  While existing 1\,mm ALMA deep fields
(Dunlop \etal\ 2016, Aravena \etal\ 2016, Franco \etal\ 2018) lack
large numbers of sources, they represent only a narrow exploration of
ALMA's blind mapping capabilities.

%
%
%

\pagebreak

\noindent \textbf{References:}

\noindent 
%
Aravena \etal\ (2016) ApJ 833, 68 \\
Aretxaga \etal\ (2011) MNRAS 415, 3831 \\
Bethermin \etal\ (2013) A\&A 557, 66\\
Bethermin \etal\ (2015) A\&A 576, 9\\
Blain \etal\ (2002) Physics Reports 369, 111\\
Casey \etal\ (2012b) ApJ 761, 140 \\
Casey \etal\ (2012c) ApJ 761, 139 \\
Casey \etal\ (2018a) ApJ 862, 77 \\
Casey \etal\ (2018b) ApJ 862, 78 \\
Casey, Narayanan \&\ Cooray (2014) Physics Reports 541, 45\\
Cowie \etal\ (2018) ApJ 865, 106\\
Daddi \etal\ (2009) ApJ 694, 1517 \\
Danielson \etal\ (2017) ApJ 840, 78 \\
De Rossi \etal\ (2018) ApJ 869, 4 \\
Dunlop \etal\ (2016) MNRAS 466, 861 \\
Dwek \etal\ (2014) ApJL 788, 30 \\
Elbaz \etal\ (2011) A\&A 533, 119 \\
Finkelstein (2016) PASA 33, 37\\
Franco \etal\ (2018) A\&A 620, 152 \\
Geach \etal\ (2017) MNRAS 465, 1789\\
Hodge \etal\ (2013) ApJ 768, 91\\
Hodge \etal\ (2016) ApJ 833, 103\\
Ivison \etal\ (2012) MNRAS 425, 1320 \\
Jiminez-Andrade \etal, in prep \\
Jones \etal\ (2013) A\&A 558, 62 \\
Magnelli \etal\ (2019) ApJ submitted \\
Maniyar \etal\ (2018) A\&A 614, 39\\
Marrone \etal\ (2018) Nature 553, 51 \\
Matsuura \etal\ (2006) MNRAS 371, 415\\
Matsuura \etal\ (2009) MNRAS 396, 918\\
Matsuura \etal\ (2011) Science 333, 1258 \\
Sanders \&\ Mirabel (1996) ARA\&A 34, 749 \\
Scott \etal\ (2008) MNRAS 385, 2225 \\
Smail \etal\ (1997) ApJL 490, 5 \\
Staguhn \etal\ (2014) ApJ 790, 77\\
Strandet \etal\ (2017) ApJL 842, 15\\
Vieira \etal\ (2013) Nature 495, 344 \\
Walter \etal\ (2012) Nature 486, 233 \\
Wang \etal\ (2011) AJ 142, 101\\
Whitaker \etal\ (2017) ApJ 850, 208 \\
Zavala \etal\ (2018) ApJ 869, 71\\

\end{document}